\documentclass[pra,twocolumn,showpacs,superscriptaddress,longbibliography]{revtex4-1}

\usepackage{graphicx}
\usepackage{dcolumn}
\usepackage{amsmath}
\usepackage{color}
\usepackage{epsfig}
\usepackage{amssymb}
\usepackage{bm}
\usepackage{color}
\usepackage{verbatim}

\newcommand{\SrCoVO}{SrCo$_2$V$_2$O$_8$}

\newcommand{\be}{\begin{equation}}
\newcommand{\ee}{\end{equation}}
\newcommand{\bea}{\begin{eqnarray}}
\newcommand{\eea}{\end{eqnarray}}
\newcommand{\up}{\uparrow}
\newcommand{\down}{\downarrow}

\def\nn{\nonumber\\}
\def\fr#1{(\ref{#1})}

\def\eps{\epsilon}

\usepackage[colorlinks,breaklinks,bookmarks=true,citecolor=blue,linkcolor=red,urlcolor=blue]{hyperref}

\begin{document}

\title{Spinon confinement in a quasi one dimensional anisotropic 
Heisenberg magnet}

\author{A. K. Bera}
\email{akbera@barc.gov.in}
\affiliation{Helmholtz-Zentrum Berlin f\"{u}r Materialien und
Energie, 14109 Berlin, Germany}
\affiliation{Solid State Physics Division, Bhabha Atomic Research Centre, Mumbai 400085, India}

\author{B. Lake}
\email{bella.lake@helmholtz-berlin.de}
\affiliation{Helmholtz-Zentrum Berlin f\"{u}r Materialien und
Energie, 14109 Berlin, Germany}
\affiliation{Institut f\"{u}r Festk\"{o}rperphysik, Technische
Universit\"{a}t Berlin, 10623 Berlin, Germany}

\author{F. H. L. Essler}
\email{Fabian.Essler@physics.ox.ac.uk}
\affiliation{The Rudolf Peierls Centre for Theoretical Physics, Oxford University, Oxford OX1 3NP, UK}

\author{L. Vanderstraeten}
\affiliation{Ghent University, Department of Physics and Astronomy,
  Krijgslaan 281-S9, B-9000 Gent, Belgium}

\author{C. Hubig}
\affiliation{Arnold Sommerfeld Center for Theoretical Physics, LMU Munich, 80333 M\"unchen, Germany}

\author{U. Schollw\"ock}
\affiliation{Arnold Sommerfeld Center for Theoretical Physics, LMU Munich, 80333 M\"unchen, Germany}

\author{A. T. M. N. Islam}
\affiliation{Helmholtz-Zentrum Berlin f\"{u}r Materialien und
Energie, 14109 Berlin, Germany}

\author{A. Schneidewind}
\affiliation{J{\"u}lich Centre for Neutron Science, Forschungszentrum J{\"u}lich GmbH,  85747 Garching, Germany}

\author{D. L. Quintero-Castro}
\affiliation{Helmholtz-Zentrum Berlin f\"{u}r Materialien und
Energie, 14109 Berlin, Germany}

\date{\today}

\begin{abstract}
Confinement is a process by which particles with “fractional” quantum numbers bind together to form quasiparticles with integer quantum numbers. The constituent particles are confined by an attractive interaction whose strength increases with increasing particle separation and as a consequence, individual particles are not found in isolation. This phenomenon is well known in particle physics where quarks are confined in baryons and mesons. An analogous phenomenon occurs in certain spatially anisotropic magnetic insulators. These can be thought of in terms of weakly coupled chains of spins $S$=1/2, and a spin flip thus carries integer spin $S$=1. Interestingly the collective excitations in these systems, called spinons, turn out to carry fractional spin quantum number $S$=1/2. Interestingly, at sufficiently low temperatures the weak coupling between chains can induce an attractive interaction between pairs of spinons that increases with their separation and thus leads to confinement. In this paper, we employ inelastic neutron scattering to investigate the spinon-confinement process in the quasi-one dimensional, spin-1/2, antiferromagnet with Heisenberg-Ising (XXZ) anisotropy \SrCoVO. A wide temperature range both above and below the long-range ordering temperature $T_N$=5.2~K is explored. Spinon excitations are observed above TN in quantitative agreement with established theory. Below $T_N$ the pairs of spinons are confined and two sequences of meson-like bound states with longitudinal and transverse polarizations are observed. Several theoretical approaches are used to explain the data. These are based on a description in terms of a one-dimensional, $S$=1/2 XXZ antiferromagnetic spin chain, where the interchain couplings are modelled by an effective staggered magnetic mean-field. A wide range of exchange anisotropies are investigated and the parameters specific to \SrCoVO\ are identified. A new theoretical technique based on Tangent-space Matrix Product States gives a very complete description of the data and provides good agreement not only with the energies of the bound modes but also with their intensities. We also successfully explained the effect of temperature on the excitations including the experimentally observed thermally induced resonance between longitudinal modes below $T_N$, and the transitions between thermally excited spinon states above $T_N$. In summary, our work establishes \SrCoVO\ as a beautiful paradigm for spinon confinement in a quasi-one dimensional quantum magnet and provides a comprehensive picture of this process.
\end{abstract}


\maketitle

\section{Introduction} 
Over the course of the last two decades,
quasi one dimensional (Q1D) quantum magnets have been established as an ideal
testing ground for key concepts of quantum many-particle physics such
as quantum criticality \cite{QC,QC2,QC3}, condensation of magnetic excitations
\cite{affleck90,tsvelik90,affleck91,condensation1,condensation2},
quantum number fractionalization \cite{FT1,FT2,KCuF3}, dimensional
crossover \cite{PhysRevLett.85.832,PhysRevB.71.134412}
and confinement of elementary particles.
Confinement originally arose in the context of high-energy physics as
a pivotal property of quarks, but subsequently was realized to emerge
quite naturally in one dimensional quantum many-particle systems and
field theories featuring kink or soliton excitations 
\cite{Mccoy.PRD.18.1259,McCoyWu.PRB.18.4886}. The
simplest such example involves domain wall (``kink'') excitations in
Ising-like ferromagnets, and has been explored in exquisite detail in
a series of experiments by Coldea and collaborators
\cite{Coldea.Science.327.177}. Confinement in ladder materials was
studied in Ref.~\onlinecite{Lake.Nat.phys.6.50}, while the confinement of spinon
excitations has been recently investigated on the Q1D spin-1/2 Heisenberg-Ising
antiferromagnetic compound BaCo$_2$V$_2$O$_8$ \cite{Grenier.PRL.114.017201}. 
Here the spinon continuum, characteristic of 1D spin-chain, observed above the three dimensional 
ordering temperature $T_N$ that breaks up into a sequence of gapped,
resolution limited modes in the 3D ordered phase ($T < T_N$). An interesting
difference to the ferromagnetic case is that two sequences of bound
states with longitudinal and transverse polarizations respectively
have been observed. 

In the present study we use inelastic neutron scattering to investigate
magnetic excitations in the Q1D spin-1/2 XXZ system \SrCoVO\ as a function of temperature covering both the 1D ($T > T_N$) and 3D ($T > T_N$) magnetic states. The
experimental results are complemented by detailed theoretical
considerations that provide a quantitative explanation of the experimental
observations.

\begin{figure} 
\centering
\includegraphics[width=0.95\linewidth,clip]{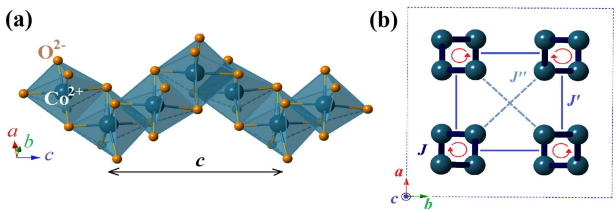}
\caption[]{\label{Fig:structure} (Color online) Crystal structure and
interactions of \SrCoVO. (a) The screw chain consisting of
edge-sharing CoO$_6$ octahedra running along the crystallographic
\emph{c}-axis. (b) Projection of the screw chains onto the
$ab$-plane. The red arrows show how the chains propagate along the
$c$-axis. The interactions between the Co$^{2+}$ ions are indicated
where $J$ is the intrachain interaction, and $J^{'}$ and $J^{''}$ are
interchain interactions.} 
\end{figure}

\SrCoVO\ crystallizes in the centrosymmetric tetragonal space group
$I4_1cd$ (No. 110) with lattice parameters $a=b=12.2710(1)$ \AA~and
$c=8.4192(1)$ \AA~at room temperature \cite{Bera.PRB.89.094402}. The
magnetic Co$^{2+}$ ions are situated within CoO$_6$ octahedra which
form edge-sharing screw chains along the crystallographic $c$-axis
\cite{Bera.PRB.89.094402,HePRB.73.212406}
(Fig.~\ref{Fig:structure}(a)). There are four screw chains per unit
cell which rotate in the $ab$-plane around (1/4,~1/4), (1/4,~3/4),
(3/4,~1/4) and (3/4,~3/4) (Fig.~\ref{Fig:structure}(b)). Two diagonal
chains rotate clockwise and the other two chains rotate anti-clockwise
while propagating along the $c$-axis. This results in a complex
interaction geometry with many possible superexchange interaction pathways. The
strongest interaction is the antiferromagnetic intrachain coupling $J$
between nearest neighboring Co$^{2+}$ ions along the chains. Weak interchain
interactions are possible along both the sides ($J^{'}$) and the
diagonals ($J^{''}$) of the $ab$-plane
(Fig.~\ref{Fig:structure}(b)). These interchain interactions are in
fact probably comprised of several interactions due to the screw chain
structure, some of which also have components along the $c$-axis as found for the isostructural compound SrNi$_2$V$_2$O$_8$ \cite{Bera.PhysRevB.91.144414}.  

The interchain interactions stabilize long-range collinear antiferromagnetic (AFM) order below $T_N = 5.2$~K \cite{Bera.PRB.89.094402} with the spins pointing parallel to the $c$-axis (chain axis).  Consecutive spins order antiferromagnetically along the chains while within the $ab$ plane, the spins order ferromagnetically/antiferromagnetically along the $a/b$ axis. The magnetic moment of the Co$^{2+}$ ions in the distorted octahedral crystal field environment is described well by a highly anisotropic \emph{pseudospin}, ${S}=1/2$ \cite{Abragram.book}. The exchange interactions between the pseudospins in \SrCoVO\ can be modeled by the Hamiltonian \cite{BonnerPR.135.A640}
\bea
\label{Eq:Ising-Heisenberg}
H &=&
J\sum_{i,j}[{S}^z_{i,j}{S}^z_{i+1,j}+\epsilon({S}^x_{i,j}{S}^x_{i+1,j}+{S}^y_{i,j}{S}^y_{i+1,j})]
\nn
&+&\sum_{i,j,n,m}J^{i,j}_{n,m}[{S}^z_{i,j}{S}^z_{n,m}+\epsilon({S}^x_{i,j}{S}^x_{n,m}+{S}^y_{i,j}{S}^y_{n,m})],\
\eea
where $S^{\alpha}_{i,j}$ is the alpha component of the $i^{th}$ spin of the $j^{th}$ chain. 
$J>0$ is AFM nearest neighbor intrachain exchange interaction and $J^{i,j}_{n,m}$ is the
interchain interaction between the $i^{th}$ spin of the $j^{th}$ chain and the $n^{th}$ 
spin of the $m^{th}$ chain. The anisotropy parameter $0<\epsilon<1$, takes into account 
the XXZ-type anisotropy interpolating between the Heisenberg ($\epsilon=1$) and Ising
($\epsilon=0$) limits. \\

\section{Experimental Methods}

Single crystals of \SrCoVO\ were grown using the floating-zone method \cite{Bera.PRB.89.094402}. Inelastic neutron scattering (INS) experiments were performed using the cold neutron triple-axis-spectrometers FLEXX at Helmholtz-Zentrum Berlin, Germany, and PANDA at the Heinz Maier-Leibnitz Zentrum, Garching,
Germany. Measurements were performed on a large cylindrical single crystal (weight $\sim 4.5$~g, diameter $\sim 4$~mm and length $\sim 40$~mm) in the ($h$,0,$l$) reciprocal space plane. The measurements
were performed with fixed final wave vectors of $k_f$=1.3~\AA$^{-1}$, $k_f$=1.57~\AA$^{-1}$ and $k_f$=1.8~\AA$^{-1}$. For these measurements, the sample was mounted on an aluminum sample holder and was cooled in cryostat. For the FLEXX spectrometer, a double focusing monochromator and a horizontally focusing analyzer were used. For the PANDA spectrometer, both monochromator and analyzer were double focusing. Higher order neutrons were filtered out by using a velocity selector on the FLEXX spectrometer and a cooled Beryllium filter on the PANDA spectrometer. Measurements took place at various temperatures between 0.8~K and 6.0~K.

\section{Experimental Results}

\subsection{High temperature phase $T>T_N$}
\label{sec:HighT}
\begin{figure}
\epsfxsize=0.5\textwidth
\includegraphics[width=0.95\linewidth]{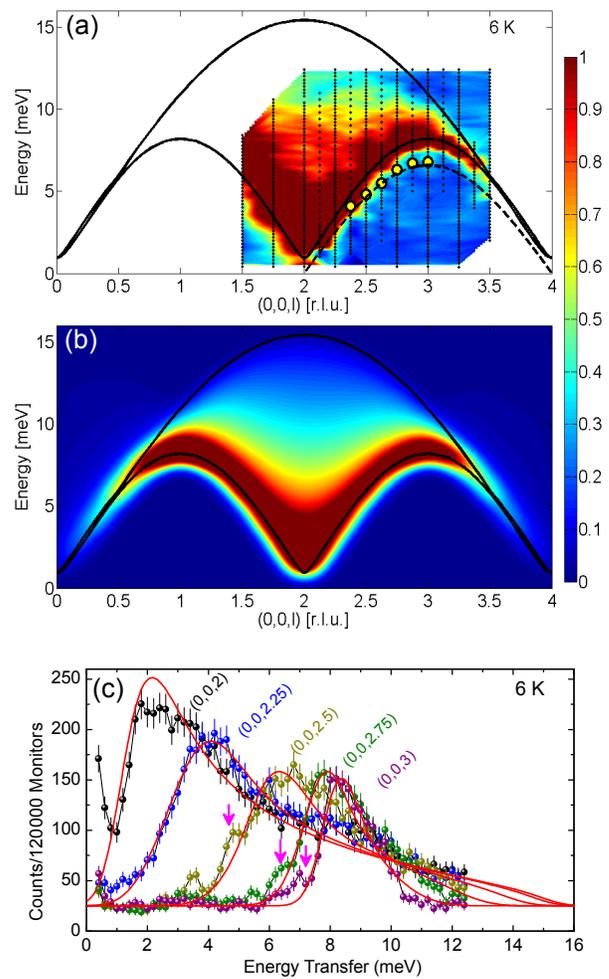}
\caption{\label{Fig:neutron_6K}(Color online) Inelastic neutron scattering from \SrCoVO\ measured in the 1D magnetic phase at 6~K ($T > T_N$), compared to theory. All the measurements were performed with $k_f$=1.57~\AA$^{-1}$. (a) Magnetic excitation spectrum along the chain axis (0,0,$l$). The spectrum was obtained by combining several energy scans performed at constant wave vector (the measured data points are represented by the black dots) and the colors indicate the size of the neutron scattering cross-section. The solid black lines are the fitted boundaries of the 2-spinon continuum of the 1D $S$ = 1/2 XXZ AFM calculated from the Bethe Ansatz (Ref. \onlinecite{Caux2008}) using $J=7.0$ meV and $\epsilon=0.56$. The dashes curve is the predicted Villain mode dispersion given by Eq. \ref{Eq:Villain_mode} and the yellow circles give the positions of the weak peaks observed below the continuum in the data. (b) The exact two-spinon contribution to the zero temperature dynamical structure factor for the 1D $S$ = 1/2 XXZ AFM spin-chain with $J=7.0$ meV and $\epsilon=0.56$ \cite{Caux2008} convolved with the instrumental resolution and multiplied by the form factor. (c) Energy scans at constant wave vectors of (0,0,2), (0,0,2.25), (0.0.2.5), (0,0,2.75) and (0,0,3) measured at 6~K. The arrows point to the observed peaks attributed to the Villain mode. The solid lines through the data are the theoretical intensities of the 1D $S$ = 1/2 XXZ AFM spin-chain convolved with the instrumental resolution and multiplied by the form factor.} 
\end{figure}

Since for \SrCoVO\ $T_N/J\ll 1$, we expect there to be a temperature regime $T_N\alt T\ll J$ in which the physics is essentially one-dimensional (1D) and approximately described by an anisotropic spin-1/2 Heisenberg XXZ chain. Single crystal inelastic neutron scattering measurements of \SrCoVO\ at 6~K ($ > T_N = 5.2$~K) along the (0, 0, $l$)-direction (chain direction) reveal a gapped scattering continuum (Fig.~\ref{Fig:neutron_6K}a). For such wave vectors the polarization factors are such that only the
components of the dynamical structure factor transverse to the direction of magnetic order contribute to the scattering cross section. The gap value of $\simeq 0.95$~meV at the (0,0,2) zone center, is quite small compared to the bandwidth of the dispersion ($\simeq 14.5$ meV) revealing that the compound lies intermediate between the Ising (gap $ \sim J$, bandwidth $ \simeq \epsilon J$) and Heisenberg (gapless, bandwidth $\simeq \pi J$) limits.  

\subsubsection{Spinon continuum at $T>T_N$}

The observed spinon continuum at $T=6$~K ($\ll J$) is in good agreement with the predictions for the transverse dynamical structure factor of the integrable Heisenberg XXZ chain at zero temperature \cite{Caux2008}. The lower boundary of the two-spinon continuum as a function of reduced momentum transfer $0\leq Q<\pi$ along the chain is given by 
\begin{eqnarray}
\label{Eq:dispersion}
\omega_l(Q)=
\begin{cases}
\omega_-(Q)
& \text{for }
0\leq Q<Q_\kappa\\ 
\frac{2I}{1+\kappa}\sin(Q) & \text{for } Q_\kappa < Q  < \pi/2\\
\omega_{\rm 1sp}(Q) & \text{for }
\pi/2<Q<\pi
\end{cases}.
\end{eqnarray}
and the upper boundary is
\begin{eqnarray}
\label{2sp_upper}
\omega_u(Q)=
\begin{cases}
\omega_{\rm 1sp}(Q) &  0 < Q  < Q_{\epsilon}\nn
\omega_-(Q) & Q_{\epsilon}<Q<\pi
\end{cases}
\end{eqnarray}
Here  
\bea
\omega_\pm(Q)&=&\frac{2I}{1+\kappa}\sqrt{1+\kappa^2\pm
  2\kappa\cos(Q)}\ ,\nn
\omega_{\rm 1sp}(Q)&=&Ik'+I\sqrt{1-(1-k'^2)\cos^2(Q)}\ ,
\eea
where $\kappa= \cos(Q_\kappa)=\frac{1-k^{'}}{1+k^{'}}$, $I = JK(\sqrt{1-k'^2})\sqrt{1-\epsilon^2}/\pi$, $K(k)$ is the complete elliptic integral of the first kind and the parameter $k'$ is given by
\be
\frac{K(k')}{K\big(\sqrt{1-k'^2}\big)}=\frac{1}{\pi}{\rm
  arccosh}\big(\epsilon^{-1}\big).
\ee
The specific value $Q_{\epsilon}$ is obtained from the solution of a quartic equation \cite{Caux2008}.

For \SrCoVO, $Q$ can be written in terms of the crystallographic wavevector transfer $Q'$  as $Q = Q'_{c} / 4 = 2 \pi l / 4$, where $Q'_{c}$ is the wave vector transfer in terms of the $c$-lattice parameter of \SrCoVO. The factor of four arises from the four equivalent Co$^{2+}$ ions per unit cell along the chain direction ($c$-axis). Fitting the experimental continuum boundaries of \SrCoVO\ to the above expressions yields the values of $J \approx 7.0 \pm 0.2 $ meV and $\epsilon \approx 0.56 \pm 0.02$. The fitted continuum boundaries are represented by the solid black lines plotted over the data in Fig.~\ref{Fig:neutron_6K}(a).

Ref.~\onlinecite{Caux2008} also provides the theoretical expression for the transverse structure factor of the 2-spinon continuum. Using the fitted values of $J$ and $\epsilon$ for \SrCoVO, the calculated transverse structure factor is shown in Fig.~\ref{Fig:neutron_6K}(b) and can be directly compared to the experimental data  in Fig.~\ref{Fig:neutron_6K}(a). Fig.~\ref{Fig:neutron_6K}(c) shows energy scans at several fixed wave vectors from (0,0,2) to (0,0,3) which pass through the lower edge of the continuum of \SrCoVO. The
lines through the data are the theoretical intensities convolved with the instrumental resolution. Good agreement is achieved between experiment and theory except at (0,0,2) where the effects of interchain coupling and finite temperature which are not included in the calculation may alter the spectrum at lowest energies. 

\subsubsection{Villain mode}

An interesting feature in the dynamical response of spin chains is the existence of a finite temperature resonance known as a Villain mode~\cite{Villain}. This ``mode'' was first observed by neutron scattering in Ref.~\onlinecite{Nagler1982,Nagler1983} and is a fairly general feature of spin chain models \cite{James.PhysRevB.78.094411,Goetze.Phys.Rev.B.82.104417}. The Villain resonance in the XXZ chain has been investigated theoretically by developing a perturbation theory around the Ising limit~\cite{James.Phys.Rev.B.79.214408}. A prediction of this theory is that above a certain temperature, a narrow resonance develops at an energy  
\be
\omega_{\rm V}(Q)\simeq {\rm max}_p|\omega_{\rm 1sp}(p)-\omega_{\rm 1sp}(p+Q)|.
\ee
The resonance corresponds to transitions between thermally occupied states and therefore disappears at zero temperature. In our case, we expect to see a resonance at low temperatures at
\be
\label{Eq:Villain_mode} 
\omega_{\rm V}(Q)\approx 0.94 J\sin(Q).
\ee
which follows a similar dispersion to that of the lower boundary of the continuum but is shifted downward from it by an energy similar to the energy gap $\simeq$ 0.95~meV. The predicted Villain mode is indicated in Fig.~\ref{Fig:neutron_6K}(a) by the dashed black curve.  As $T=6$K which is still quite low compared to the intrachain interaction ($T/J\approx 0.07$), we expect the temperature effects on the $T=0$  two-spinon continuum to be weak. Hence, the most noticeable effect of temperature is the emergence of additional peaks associated with the Villain mode in the $T=6$~K data just below the two-spinon continuum. A weak peak is indeed visible in the (0,0,3) data at $\approx 6.8$~meV and in the (0,0,2.75) and (0,0,2.5) scans at $\approx 6.3$~meV and $\approx 4.8$~meV, respectively (see Fig.~\ref{Fig:neutron_6K}(c)). These peak positions along with those obtained from other energy scans (not shown) are represented by the yellow circles in Fig.~\ref{Fig:neutron_6K}(a) and follow the predicted Villain mode dispersion given by the dashed black curve. 

\subsection{Low temperature phase $T < T_N$}
\label{ssec:lowTP}

\begin{figure}
\centering
\epsfxsize=0.5\textwidth
\includegraphics[width=0.95\linewidth,clip]{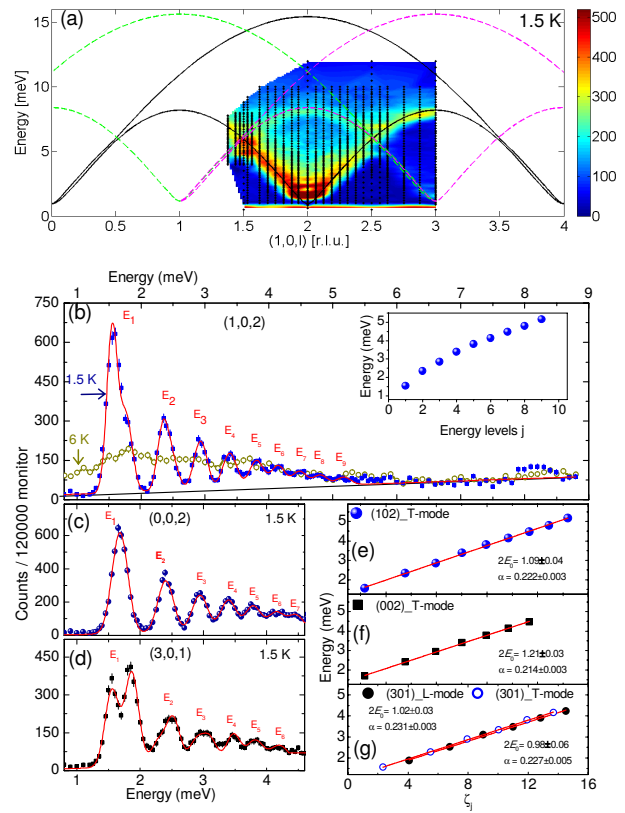}
\caption[]{\label{Fig:Neutron_1p5K} 
(Color online) Neutron scattering data in the ordered phase at $T=1.5$~K. (a) Scattering intensity along the 
chain direction at (1,0,$l$) measured with $k_f$=1.8~\AA$^{-1}$. The different colored lines delineate the boundaries of the independent and overlapping spectra expected for uncoupled chains. (b) Scans in energy at (1,0,2) for $T = 1.5$~K ($< T_N = 5.2$~K) and $T = 6$~K ($> T_N$) with $k_f$=1.57~\AA$^{-1}$. The red curve is a fit of Gaussian peaks to the data. The background is shown by the black line. Inset: energies of the transverse bound spinon mode excitations as a function of mode number. (c)-(d) Energy scans at constant wave vectors (0,0,2) and (3,0,1) with $k_f$=1.57~\AA$^{-1}$. Only the T-mode is observed at (0,0,2), while both T-mode and L-mode are seen at (3,0,1). The solid red curves are fits of Gaussian peaks to the data. (e)-(g) Fits of the observed bound state energies to a model of two spinons interacting with an attractive interaction increasing linearly with their separation, \emph{cf.} \ref{Schroedinger}. The energies of a given series of modes (L or T) are plotted against the negative zeros of the Airy function $\zeta_j$. The solid red lines are the linear fits to the data, demonstrating linear confinement. The fitted values of $E_0$ and $\alpha$ are given on the plots.} 
\end{figure}

Below its N\'{e}el temperature $T_N=5.2$~K, \SrCoVO\ develops long-range magnetic order where the Co$^{2+}$ spins order antiferromagnetically along the chains with their moments aligned parallel to the $c$-axis \cite{Bera.PRB.89.094402}. The dynamical structure factor well inside the ordered phase at $T=1.5$~K along the (1,0,$l$) direction is shown in Fig.~\ref{Fig:Neutron_1p5K}. Its gross features including the total bandwidth and the energy gap are similar to those observed above $T_N$ (Fig.~\ref{Fig:neutron_6K}(a)). The weak scattering at (1,0,3) is associated with the fact that there
are four equivalent screw chains per unit cell each with four Co$^{2+}$ ions per $c$-lattice parameter. Neglecting interchain interactions this gives rise to a total of four ``copies'' of the cross section for a single chain, which are shifted with respect to one another by reciprocal lattice units along the chain direction (for details see Ref. \onlinecite{Bera.PhysRevB.91.144414} on the isostructural compound SrNi$_2$V$_2$O$_8$). For uncoupled chains we thus expect the intensity to be of the form
\be
I({{\bf Q}'},\omega)=\sum_{l=1}^4 A_l({\bf Q}') I_{\rm 1D}\left(4Q+\frac{2\pi(l-1)}{c},\omega\right)\ 
\ee
As a result every reciprocal lattice point is an antiferromagnetic zone center for at least one of these copies, but their overall intensities $A_l({\bf Q}')$ depend on the full momentum transfer ${\bf Q}'$ and can be very different. For (1,0,$l$) all four independent contributions are present shifted consecutively by $\Delta Q_{c}'=1$ r.l.u. along the chain. Their lower and upper boundaries are indicated by the different colored lines in Fig.~\ref{Fig:Neutron_1p5K}(a). For (0,0,$l$) only a single contribution is visible, as observed in Fig.~\ref{Fig:neutron_6K}(a).   

Careful inspection of the cross section at the antiferromagnetic zone center, reveals that the continuum observed at 6~K is transformed into a sequence of discrete, resolution-limited excitations at 1.5~K. As shown in Fig.~\ref{Fig:Neutron_1p5K}(b) at wave vector transfer (1,0,2) nine peaks, labelled $E_1$-$E_9$, are observable in the energy range $\sim$~1.5-5.5~meV. Since these discrete modes appear below the ordering temperature, they must arise from the interchain coupling. A detailed examination shows that each of the sharp peaks at (1,0,2) in fact consists of two closely spaced peaks with the higher energy peak being relatively weaker. For the wave vector (0,0,2) a single series of peaks is found (Fig.~\ref{Fig:Neutron_1p5K}(c)), while at (3,0,1) both series of peaks are visible with similar intensities (Fig.~\ref{Fig:Neutron_1p5K}(d)).  

\begin{figure}
\centering
\epsfxsize=0.5\textwidth
\includegraphics[trim=1cm 0cm 0cm 6cm, width=0.95\linewidth,clip]{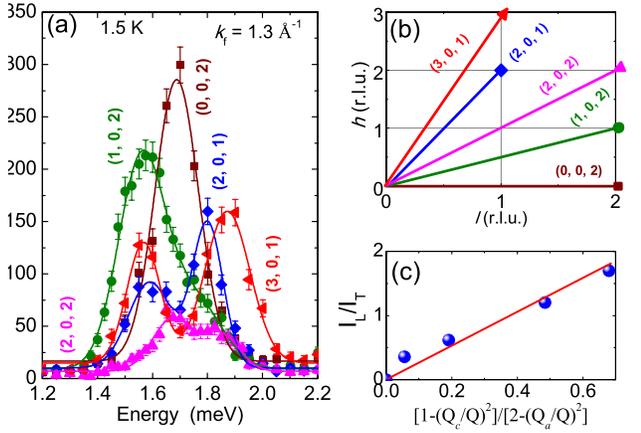}
\caption[]{\label{Fig:L-T_modes} (Color online) (a) Constant wave vector scans for several reciprocal lattice points over the lowest pair of bound spinon modes at $\sim 1.5$~meV showing how the L-mode and T-mode intensities vary with wave vector in the ($h$,0,$l$)-plane. (b) Directions of the scattering
wave vectors in the reciprocal plane for the scans shown in  (a). (c) Intensity ratio of the L-mode to the T-mode  as a function of $[1-(Q_c'/Q')^2]/[2-(Q_a'/Q')^2]$. The red curve is a linear fit to the data.} 
\end{figure}

We have investigated the nature of the two series of peaks in more detail at several AFM zone centers with different wave vector components $Q_a'$ and $Q_c'$ (along the $a$ and $c$ axes) respectively. The measurements were performed over the lowest energy peaks around $\sim 1.5$~meV, see Fig.~\ref{Fig:L-T_modes}(a). The results indicate that when the wave vector transfer is parallel to the chain direction, e.g. (0,0,2), only one peak is present. If the $a$-component of the wave vector transfer is non-vanishing a second peak appears at higher energy. The relative intensity of the higher energy peak increases with increasing $Q_a'$. 
 
This intensity dependence provides important information about the nature of the two series of peaks. Neutron scattering is only sensitive to fluctuations perpendicular to the wave vector transfer. The higher energy series of modes that is absent for wave vector transfers parallel to the $c$-axis but becomes visible when $Q_a'\neq 0$ must therefore be due to fluctuations along to the $c$-axis. We refer to this series of modes as \emph{longitudinal modes} (L-modes) since they are due to fluctuations parallel to the ordered spin direction. In contrast, the available evidence suggests that the lower energy series of modes is associated with fluctuations in the $ab$-plane. We will therefore refer to these excitations as \emph{transverse modes} (T-modes). The bound modes in \SrCoVO\ were observed previously using terahertz spectroscopy as described in Ref. \onlinecite{Zhe.PRB.91.140404}. This techniques allows the transverse modes to be measured to very high resolution, but the longitudinal modes are not accessible.

\begin{figure}
\centering
\includegraphics[width=0.95\linewidth,clip]{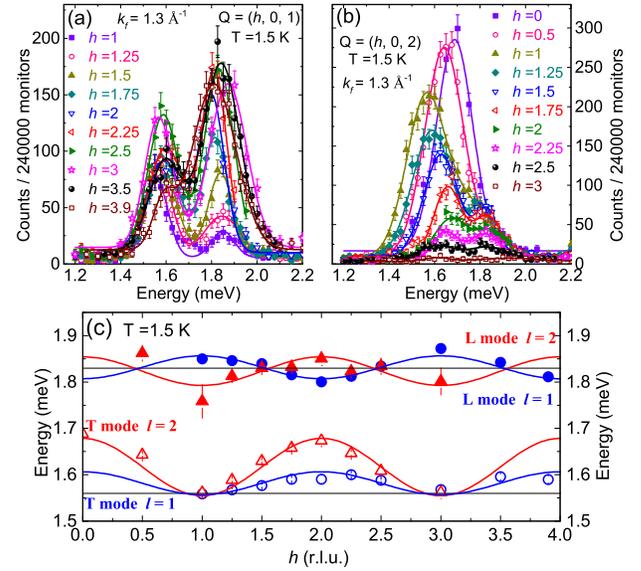}
\caption[]{\label{Fig:perp_dispersion} (Color online) Constant wave vector scans at 1.5~K for (a) ($h$, 0, 1) and (b)   ($h$, 0, 2). The energy window covers the lowest pair of bound-spinon modes and yields the dispersions perpendicular to the chain axis. The solid lines are fits by sums of two Gaussians and allow us to extract the energies of the T- and L-modes. (c) The dispersion relations perpendicular to the chain axis for the T-mode and L-mode constructed from the extracted peak positions from (a) and (b). The
solid lines are the guides to the eye.} 
\end{figure}

It is clear from Fig.~\ref{Fig:L-T_modes} that the energies of the modes vary from one AFM zone center to another as a result of the interchain interactions. In order to investigate these interactions, the dispersions of the lowest energy pair of bound modes were measured along $Q'_a$ by performing a series of energy scans at the constant-wave vectors ($h$,~0,~1) and ($h$,~0,~2) for various values of $h$ (Figs.~\ref{Fig:perp_dispersion}(a) and \ref{Fig:perp_dispersion}(b)). Both the L-mode and T-mode disperse over a narrow bandwidth of $\le 0.15$~meV, the modes are in-phase for $l$=2 but out-of-phase at $l$=1 (Fig.~\ref{Fig:perp_dispersion}(c)). The dispersions are complex due to the many possible interchain interactions allowed by the screw-chain crystal structure (Fig.~\ref{Fig:structure}) that can reinforce or act against each other depending on the reciprocal lattice points as found for the isostructural compound
SrNi$_2$V$_2$O$_8$ \cite{Bera.PhysRevB.91.144414}. To fully quantify the strengths of the interchain interactions further measurements are required. 

\subsubsection{Modeling the energies of the observed bound modes}
\label{Schroedinger}
As we will detail in section \ref{sec:theory}, the bound states observed at low temperatures can be understood in terms of \emph{confinement of spinon pairs}. The physical picture is that the interchain coupling induces a linearly confining potential between the elementary spinon excitations of the 1D chains. This was shown by Shiba in Ref.~\onlinecite{shiba} in the large anisotropy limit $\epsilon\ll 1$ of the model 
\fr{Eq:Ising-Heisenberg}. In this limit spinons can be thought of as antiferromagnetic domain walls. As we will see in section \ref{sec:theory} the spinon confinement picture extends all the way up to the Heisenberg limit $\epsilon=1$. This suggests that the bound mode energies $E_j$ at the AFM zone center can be
approximately extracted from the 1D Schr\"{o}dinger equation describing the centre-of-mass motion of the spinon pairs  

\begin{equation}
\label{Eq:SpinnonSchroedingerEq}
-\frac{\hbar^2}{\mu}\frac{d^2\varphi}{dx^2}+\lambda |x|
\varphi=(E-2E_0)\varphi\ .
\end{equation}

Here $\mu$ is the reduced mass, $E_0$ is the spinon gap in absence of the confining potential, $\lambda$ is the molecular field at the Co$^{2+}$ site produced by the interchain interactions, and the interaction potential between the two spinons is assumed to be a linear function of their separation $x$. The Schr\"odinger equation \fr{Eq:SpinnonSchroedingerEq} has been previously applied successfully to describe aspects of confinement in the transverse field Ising chain \cite{Mccoy.PRD.18.1259,McCoyWu.PRB.18.4886,Isingconfinement1,BT,Rutkevich2010} and in real materials
\cite{Coldea.Science.327.177,Grenier.PRL.114.017201}. The solutions of Eq.~(\ref{Eq:SpinnonSchroedingerEq}) are given by Airy functions~\cite{Landau} and the corresponding bound state energies are

\begin{equation}
\label{Eq:SpinnonSchroedingerSolution}
E_j=2E_0+\alpha \zeta_j \hspace{5mm}  j= 1, 2, 3, ....,
\end{equation}

where $\alpha = [\lambda^2 (\hbar^{2}/\mu)]^{1/3}$ and the $\zeta_j$'s are the negative zeros of the Airy function. We use Eq.~\fr{Eq:SpinnonSchroedingerSolution} as a phenomenological expression for the bound state energies, and fit the two parameters $E_0$ and $\alpha$ to our experimental data for the longitudinal and transverse modes separately. This gives excellent agreement with the observed
spectra in all cases see Fig.~\ref{Fig:Neutron_1p5K}(e) to (g). The fitted value of $\alpha$ is $\alpha=0.22 \pm 0.1$ while the spinon gap $0.98 < 2E_0 < 1.21$ shows some variation between different AFM zone centers probably due to the interchain coupling. 

It should be noted that the bound modes have their minimum at the reciprocal lattice points and disperse along the chain direction as can be observed in Fig.~\ref{Fig:Neutron_1p5K}(a). The bound mode dispersion in the vicinity of the antiferromagnetic zone center is of the form 
\be
\Omega^{(a)}_j(Q)\approx E^{(a)}_j+\frac{(Q-\pi)^2}{2m^{(a)}_j},\quad a=L,T,
\ee
where $Q$ is the reduced wave vector along the chain direction ($Q=Q'_{c}/4$) and $m_j$ is the mass of the $j^{\rm th}$ bound state.

\subsection{Temperature effects}
\label{ssec:T>0}
\begin{figure}
\begin{center}
\includegraphics[width=0.95\linewidth]{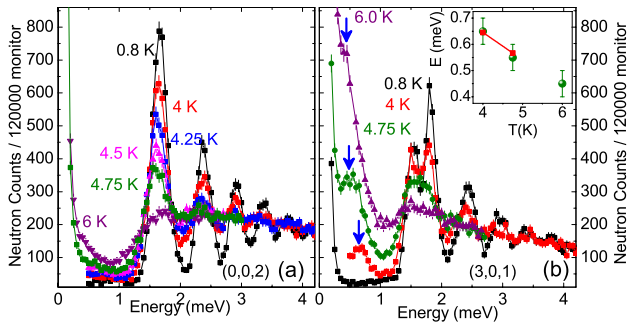}
\end{center}
\caption{\label{Fig:temp_boundstates}(Color online) Energy scans at (a) (0,0,2) and (b) (3,0,1), measured with $k_f$=1.57~\AA$^{-1}$, for several temperatures close to $T_N$. Temperature induced peaks appear for (3,0,1) and are indicated by arrows. The inset shows the energy of the temperature induced peak as a function of temperature (green circles). The energy difference between first two longitudinal bound states is
shown by the red squares. } 
\end{figure}

In the simplest model the confining potential for spinons is proportional to the magnitude of the ordered moment and we therefore expect the bound modes to be sensitive to temperature at $T\approx T_N$. The
temperature-dependence of the transverse and longitudinal bound spinon modes at the reciprocal lattice points (0,0,2) and (3,0,1) are shown in Fig.~\ref{Fig:temp_boundstates}. As temperature approaches $T_N$ from below these modes broaden, become weaker and shift to lower energy. This shift is due to the weakening of the confining molecular field from the neighboring chains as the order moment value decreases with increasing temperature. 

Another feature in the data is a strong broad peak centered around $E=0$ [Fig.~\ref{Fig:temp_boundstates}(b)]. It is visible at $T \sim T_N$ but disappears for $T \ll T_N$ suggesting that it is due to short-range order between the chains that sharpens into magnetic Bragg peak position well below $T_N$. We observe that the peak is present at (3,0,1) but not at (0,0,2). The likely origin of this difference is that at (0,0,2) we only observe transverse correlations, while the peak is related to (emerging) 3D order along the longitudinal direction.

In addition to these changes, a sharp peak appears at (3,0,1) at the energy $E \sim 0.65$~meV for $T=4$~K and shifts towards lower energy with increasing temperature. No such temperature-induced peak is observed at (0,0,2), confirming that this feature is associated with the longitudinal structure factor. We attribute this peak to transitions between bound modes. At finite temperatures the lowest energy bound
mode will become thermally populated and transitions between it and the higher energy bound modes are possible. Since the peak is observed in the longitudinal structure factor it arises from a transition between longitudinal bound modes or between transverse bound modes but not from a longitudinal to a transverse bound mode or vice versa. The distinction should be drawn between this feature which is a transition between thermally excited bound-spinon modes observed close to but below $T_N$ and the Villain mode which arises from transitions between thermally excited single spinon states which are observed above $T_N$.

If we denote the dispersion relation of the j$^{th}$ transverse/longitudinal bound state by $\Omega^{(a)}_j(Q)$ ($a=T,L$), transitions between them occur at energies
\be
\Omega_{j,k}^{(a\rightarrow
  b)}(Q-P)=|\Omega^{(a)}_j(Q)-\Omega^{(b)}_j(P)|\ ,\quad a,b=T,L.
\label{omeganm}
\ee
In the low temperature ordered phase $\Omega^{(a)}_j(Q)$ will be $\pi$-periodic functions, so that the momentum transfer of the transition \fr{omeganm} will be $Q-P\ {\rm mod}\ \pi$. If we take $Q=P=\pi$ there will be transitions with momentum transfer zero and $\pi$. Setting aside the issue which transitions will give non-vanishing contributions to the dynamical structure factor (which is $2\pi$-periodic), we have checked whether the energy differences between bound modes at (3,0,1) are reflected in the energy
of the temperature induced peaks. The inset of Fig.~\ref{Fig:temp_boundstates}(b) shows the temperature-dependent energy of the thermally excited peak (green circles) which is in good agreement with the energy difference between the first and second longitudinal bound modes as a function of temperature (red squares). The above interpretation of temperature effects is supported by the theoretical analysis summarized in section \ref{ssec:tempeff}. 

\section{Theory}
\label{sec:theory}
As we have seen above, at $T>T_N$ the neutron scattering intensity is
well described by a model of uncoupled spin-1/2 Heisenberg XXZ chains.
At low temperatures interchain coupling effects are obviously
important. In the following we constrain our analysis to a simple
mean-field treatment of these interactions\cite{shiba,MFA1,MFA2,MFA3,review}
\bea
\sum_{i,j,n,m}J^{i,j}_{n,m}[{S}^z_{i,j}{S}^z_{n,m}+\epsilon({S}^x_{i,j}{S}^x_{n,m}+{S}^y_{i,j}{S}^y_{n,m})]\nn
\longrightarrow
\sum_{i,j,n,m}J^{i,j}_{n,m}\left[\langle{S}^z_{i,j}\rangle{S}^z_{n,m}
+{S}^z_{i,j}\langle{S}^z_{n,m}\rangle\right].
\eea
Using the fact that there is N\'eel order at low temperatures this
leads to a description in terms of decoupled chains in a
self-consistent staggered magnetic field 
\bea
H_{\rm MF}&=&J\sum_{j=1}^L
S^z_jS^z_{j+1}+\epsilon\big(S^x_jS^x_{j+1}+S^y_jS^y_{j+1}\big)\nn
&&-h\sum_{j=1}^L(-1)^jS^z_j \ .
\label{HMF}
\eea
The effective staggered field $h$ is a function of $J^{i,j}_{n,m}$ and
temperature. We note that the Hamiltonian \fr{HMF} has a U(1) symmetry
of rotations around the z-axis
\be
[H_{\rm MF},S^z]=0.
\label{U1}
\ee
In the following we analyze the dynamical structure factor in the
model \fr{HMF} by several different methods in various parameter
regimes. 

\subsection{Strong coupling expansion}
A fairly comprehensive qualitative picture of the physical properties
of the model \fr{HMF} can be obtained by considering the strong
anisotropy limit $\epsilon\ll 1$. This limit is amenable to an
analysis by the method of Ishimura and Shiba~\cite{IS} and has been
previously considered by Shiba~\cite{shiba}. As
Ref.~\onlinecite{shiba} only considered the transverse component of
the dynamical structure factor, we now give a self-contained
discussion of this approach and then discuss the resulting picture for
dynamical correlations. 
We find it convenient to map \fr{HMF} to a ferromagnet by rotating
the spin quantization axis on all odd sites around the x-axis by 180
degrees  
\be
S^a_{2j+1}=-\frac{1}{2}\tau^a_{2j+1}\ ,\ a=y,z\ ,\quad
S^x_{2j+1}=\frac{1}{2}\tau^x_{2j+1}.
\ee
Here $\tau_j^\alpha$ are Pauli matrices. In terms of the new spins we
have $H_{\rm MF}=H_0+H'$ with
\bea
H_0&=&-\frac{J}{4}\sum_j \tau^z_j\tau^z_{j+1}\ ,\nn
H'&=&\frac{J\epsilon}{2}\sum_j\tau^+_j\tau^+_{j+1}+\tau^-_j\tau^-_{j+1}
-\frac{h}{2}\sum_j\tau^z_j,
\label{HMF2}
\eea
where $\tau^\pm=\frac{\tau^x\pm i\tau^y}{2}$. The U(1) symmetry
\fr{U1} gives rise to the commutation relations
\be
[H_{\rm MF},\sum_{j}(-1)^j\tau^z_j]=0.
\label{U1b}
\ee
The zero temperature dynamical susceptibilities are given by
\bea
\chi^{\alpha\beta}(\omega,Q)&=&-\frac{i}{L}\int_0^\infty dt\ 
\sum_{j,l}e^{i\omega t+iQ(j-l)}
\langle[\tau^\alpha_l(t),\tau^\beta_j]\rangle
\nn
&=&-\langle{\rm
  GS}|\tau^\beta_Q\frac{1}{\omega+H_{\rm MF}-E_0+i\eta}\tau^\alpha_{-Q}|{\rm
  GS}\rangle\nn
&+&\langle{\rm
  GS}|\tau^\alpha_{-Q}\frac{1}{\omega-H_{\rm
    MF}+E_0+i\eta}\tau^\beta_{Q}|{\rm GS}\rangle,\
\label{chialphabeta}
\eea
where $\eta>0$ is infinitesimal, $E_0$ is the ground state energy and
$
\tau^\alpha_Q=\frac{1}{\sqrt{L}}\sum_j e^{iQj}\tau^\alpha_j.
$
The dynamical structure factor of the antiferromagnetic spin chain
\fr{HMF} of interest is
\bea
S^{xx}_{\rm AFM}(\omega,Q)&=&S^{yy}_{\rm AFM}(\omega,Q)
=-\frac{1}{4\pi}{\rm
  Im}\ \chi^{xx}(\omega,Q)\ ,\nn
S^{zz}_{\rm AFM}(\omega,Q)&=&-\frac{1}{4\pi}{\rm
  Im}\ \chi^{zz}(\omega,Q+\pi)\ .
\label{DSF}
\eea
We will analyze \fr{chialphabeta} by carrying out a strong coupling expansion
in the limit $\epsilon,h\ll J$~\cite{IS}. Our starting 
point is the Ising part $H_0$ of the mean field Hamiltonian. The
ground states of $H_0$ are simply the saturated ferromagnetic states
$|\up\rangle$ and $|\down\rangle$ respectively. Their energies are 
$
E^{(0)}_0=-\frac{JL}{4}.
$
Spontaneous symmetry breaking selects e.g. $|\up\rangle$. The low-lying
excitations are then 2-domain wall states of the form
$
...\up\up\up\down\down\down\up\up\up...
$
We denote these by $|j,n\rangle$ where $j$ is the position of the
first down spin and $j+n-1$ the position of the last down spin.
The energies of these states are
$
E^{(0)}_2=-\frac{JL}{4}+J.
$
A convenient orthonormal basis of 2-domain wall states with momentum
$Q$ is obtained by taking appropriate linear combinations
\be
|Q,n\rangle=\frac{1}{\sqrt{L}}\sum_je^{iQj}|j,n\rangle.
\ee
The matrix elements of the Hamiltonian in these states are
\bea
\langle Q,m|H|Q,n\rangle&=&\frac{J\epsilon}{2}(1+e^{2iQ})\left[\delta_{m,n+2}
+e^{-2iQ}\delta_{m,n-2}\right]\nn
&+&
\left[\big(-\frac{J}{4}-\frac{h}{2}\big)L
+J+hn\right]\delta_{n,m}.
\label{matrixelements}
\eea
Importantly, the only non-zero matrix elements in $H$ occur between
domain-wall states with both even or both odd lengths. This is a
consequence of the U(1) symmetry \fr{U1b} and expresses the fact that
acting with $H$ does not change the staggered magnetization (or
equivalently the magnetization in the original $S^\alpha_j$ spin
variables). Given \fr{matrixelements} it is a straightforward matter
to numerically compute the Green's functions 
\be
G(j,k)=\langle Q,j|\left(\omega-H_{\rm MF}+E_0+i\eta\right)^{-1}|Q,k\rangle.
\ee

\subsubsection{Ground state in perturbation theory}
The first order correction to the ground state is obtained by standard
perturbation theory
\bea
|{\rm GS}\rangle&\simeq&
|\up\rangle-\frac{\epsilon\sqrt{L}}{2}|Q=0,2\rangle.
\eea
This gives the following matrix elements of spin operators
\bea
\langle Q,n|\tau^x_Q|{\rm GS}\rangle&=&
\left(1-\epsilon\cos(Q)\right)\delta_{n,1}
-\frac{\epsilon(1+e^{2iQ})}{2}\delta_{n,3}\ ,\nn
\langle Q,n|\tau^z_Q|{\rm GS}\rangle&=&
\epsilon(1+e^{iQ})\delta_{n,2}\delta_{Q,0}\ ,\nn
\langle{\rm GS}|\tau^z_Q|{\rm GS}\rangle&=&\sqrt{L}\delta_{Q,0}\ .
\label{ME2}
\eea
\subsubsection{Dynamical Structure Factor}
\begin{figure}
\begin{center}
\includegraphics[width=0.75\linewidth]{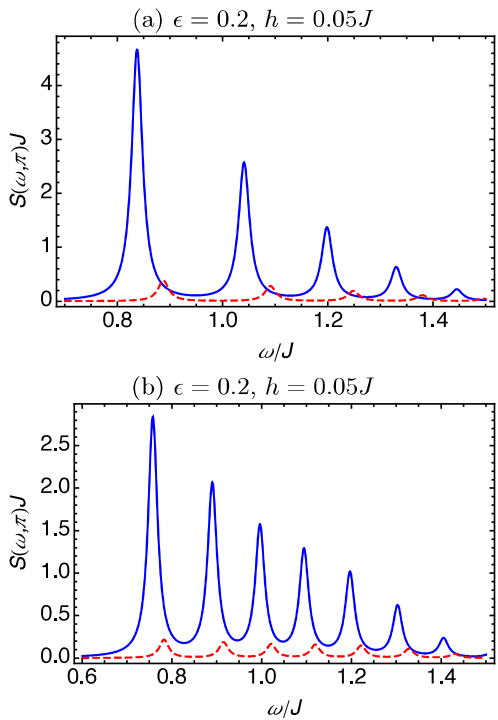}
\end{center}
\caption{(Color online) Transverse DSF $S^{xx}_{\rm AFM}(\omega,\pi)$ (solid blue line) and longitudinal DSF $S^{zz}_{\rm AFM}(\omega,\pi)$ (dashed red line) for (a) $\epsilon=0.2$ and $h=0.05J$, and (b) $\epsilon=0.2$ and $h=0.025J$. The broadening has been chosen as $\eta=J/80$ in order to make the delta-function peaks visible.} 
\label{fig:both_pi}
\end{figure}

Substituting \fr{ME2} into \fr{chialphabeta} then leads to the
following approximate expression for the dynamical susceptibilities
at $\omega>0$ 
\bea
\chi^{xx}(\omega,Q)&\simeq&G(1,1)
\left(1-\epsilon\cos(Q)\right)^2
+G(3,3)
|\epsilon\cos(Q)|^2\nn
&-&G(1,3)
\left(1-\epsilon\cos(Q)\right)
\frac{\epsilon(1+e^{2iQ})}{2}\nn
&-&G(3,1)
\left(1-\epsilon\cos(Q)\right)
\frac{\epsilon(1+e^{-2iQ})}{2}
\label{sxxfull}\nn
\chi^{yy}(\omega,Q)&=&
\chi^{xx}(\omega,Q+\pi)\ ,
\label{syyfull}\nn
\chi^{zz}(\omega,Q)&\simeq&G(2,2)
\left|2\epsilon\cos(Q/2)\right|^2.
\eea
We note that these are consistent with the U(1) symmetry of rotations
around the z-axis for the antiferromagnetic model \fr{HMF}.
It is now straightforward to compute the dynamical structure factor (DSF)
\fr{DSF} numerically. Results for momentum transfer $\pi$ are shown in
Figs.~\ref{fig:both_pi}.

\begin{figure}[ht]
\includegraphics[trim=0cm 11cm 0cm 2.5cm, width=0.95\linewidth,clip]{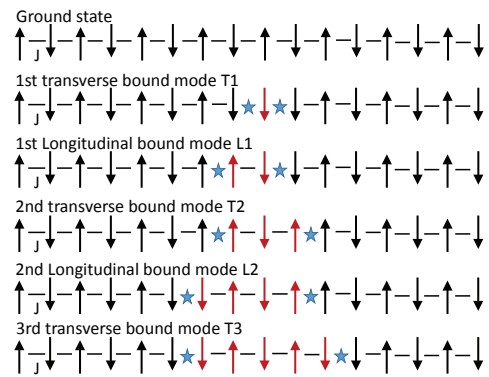}
\caption{\label{Fig:boundstates}(Color online) 
Cartoon picture of the spin excitations for $\epsilon\ll 1$. (a) 
Lowest energy configuration showing antiferromagnetic 
spin alignment. (b) Flipping one spin (red arrow) creates two
ferromagnetic domain walls (blue stars). At $T>T_N$
these ``spinons'' can propagate independently and are observed in the
INS data as a continuum. At $T<T_N$ the domain walls are confined
by the molecular field due to the ordering of the neighboring
chains: the energy cost increases linearly with the number of reversed
spins (red arrows), which leads to the hierarchy of bound modes
observed in the INS data. Domain walls separated by an even (odd) number of 
flipped spins have $S^z=0$ ($S^z=1$) and are observed in the
longitudinal (transverse) structure factor.
}
\end{figure}

We see that the transverse DSF ($S^{xx}_{\rm AFM}(\omega,\pi)$) only `couples' to half the bound
states, while the longitudinal DSF ($S^{zz}_{\rm AFM}(\omega,\pi)$) is sensitive to the other
half. This is in perfect correspondence with the experimental
observations. The selection rule that gives rise to this behaviour is
related to the conserved $S^z$ quantum number \fr{U1}. 
It is clear from \fr{matrixelements} that the Hamiltonian in the
2-domain wall sector is block diagonal in a basis of domain wall
states of odd/even length. In terms of the original spins even/odd
length domain walls correspond to even/odd values of the conserved
$S^z$ quantum number (assuming the lattice length to be divisible by
4). This implies that there is one sequence of bound states with
$S^z=0$, and a second with $S^z=\pm 1$. The first is visible in the
longitudinal structure factor $S^{zz}_{\rm AFM}$, while the second
contributes only to $S^{xx}_{\rm   AFM}$.
In the strong anisotropy limit we therefore have the simple cartoon
picture for the physical nature of the bound states shown in
Fig.~\ref{Fig:boundstates}.

\subsubsection{Gap as a function of field}
\begin{figure}[ht]
\begin{center}
\epsfxsize=0.43\textwidth
\includegraphics[width=0.95\linewidth]{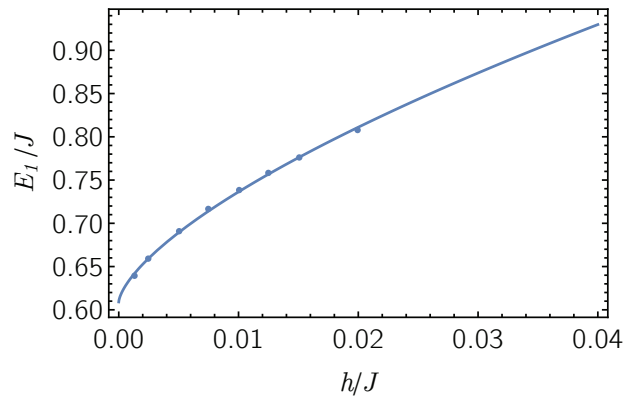}
\end{center}
\caption{(Color online) Gap between the ground state and the lowest excitation at $Q=\pi$ as a function of the staggered field $h$ for $\epsilon=0.2$. Dots are results of the strong coupling expansion while the solid line is a fit to equation \fr{gapfit}.}
\label{fig:gap}
\end{figure}

The position of the first peak at $Q=\pi$, $E^{(T)}_1$, gives the excitation gap. Based on the relation of our problem to a Schr\"odinger equation with linear potential we expect
\be
E^{(T)}_1=a_0+a_1h^\frac{2}{3}.
\label{gapfit}
\ee
In Fig.~\ref{fig:gap} we show the results obtained in our strong coupling expansion and a fit to \fr{gapfit}, which is seen to give a very good account of the data [see Fig. \ref{fig:dmrg1}]. 

\subsection{Field Theory in the vicinity of the isotropic point $\epsilon=1$}
The physical picture obtained in the large anisotropy limit
$\epsilon\ll 1$ remains valid in the entire regime
$0<\epsilon<1$. To see this we consider the limit of weak
anisotropy $\epsilon\approx 1$, where the mean-field Hamiltonian
\fr{HMF} can be written in the form 
\be
H_{\rm latt}=J\sum_j {\bm S}_j\cdot{\bm S}_{j+1}
+\delta\sum_jS^z_jS^z_{j+1}-h\sum_j(-1)^jS^z_j\ 
\ee
where $\delta \approx J(1-\epsilon)$. In the parameter regime $h\ll\delta\ll J$ 
this model can be bosonized following e.g. \cite{boso,boso2}, which leads to 
a two-frequency sine-Gordon model 
\bea
{\cal H}&=&\frac{v}{2}\int dx\left[\big(\partial_x\Phi\big)^2+
\big(\partial_x\Theta\big)^2\right]\nn
&+&\int dx\left[\lambda\cos\sqrt{8\pi}\Phi(x)
+\mu\sin\sqrt{2\pi}\Phi(x)\right] ,
\label{DSG}
\eea
where $\lambda\propto\delta$ and $\mu\propto h$. The model \fr{DSG}
in the relevant parameter regime has been studied previously by a
number of authors \cite{affleck,DM,BT}. A fruitful line of attack is to
start with ${\cal H}$ for $\mu=0$, and consider the $\mu$-term as a
perturbation. The minima of the potential for $\mu=0$ and $\lambda>0$
occur at 
\be
\Phi(x)=(2n+1)\sqrt{\frac{\pi}{8}}\ ,\quad n\in\mathbb{Z}.
\ee
The solutions to the classical equations of motion are solitons and
anti-solitons. Solitons interpolate between neighbouring vacua, e.g.
\be
\Phi(x\to \pm\infty)=\mp\sqrt{\frac{\pi}{8}},
\ee
while antisolitons have the opposite asymptotics
\be
\Phi(x\to \pm\infty)=\pm\sqrt{\frac{\pi}{8}}.
\ee
At the quantum level solitons and antisolitons turn into elementary
excitations of the sine-Gordon model.
\subsubsection{Soliton-antisoliton states}
Following \cite{affleck,DM} we start with the soliton-antisoliton
sector. We take the positions of the soliton and antisoliton to be
$x_1$ and $x_2$ respectively and denote the classical energy for
$\mu=0$ by $2\Delta_s$. When $\mu>0$, soliton-antisoliton states
acquire an extra contribution to the energy
\be
2\mu|x_1-x_2|.
\label{Vconf}
\ee
In a non-relativistic approximation we then obtain a single-particle
Schr\"odinger equation for the relative motion ($x=x_2-x_1$) with
Hamiltonian 
\be
H_{\rm rel}=-\frac{1}{M_s}\frac{d^2}{dx^2}+2\mu |x|\ .
\label{SG2}
\ee
Here $M_s=\Delta_s/v^2$ and the reduced mass is
$\frac{M_sM_{\bar s}}{M_s+M_{\bar s}}=\frac{M_s}{2}$.
This Schr\"odinger equation can be solved exactly in terms of
Airy functions \cite{Landau}, and the corresponding eigenstates
describe the confinement on solitons and antisolitons. The
bound state energies follow from the boundary conditions imposed on
the wave function. If we require the wave function to be antisymmetric
and therefore vanish at zero, we obtain
\be
E^{(L)}_n=2M_s+\left(\frac{4\mu^2}{M_s}\right)^\frac{1}{3}\xi_n\ ,\quad {\rm Ai}(-\xi_n)=0.
\label{Elong}
\ee
Symmetric wave functions would instead lead to a
spectrum of the form
\be
E^{(L)}_n=2M_s+\left(\frac{4\mu^2}{M_s}\right)^\frac{1}{3}\zeta_n\ ,\quad {\rm Ai}'(-\zeta_n)=0.
\label{Elong2}
\ee
As soliton-antisoliton states have the same $S^z$ value as the ground
state, the bound states \fr{Elong} will be visible in the longitudinal
structure factor. 
\subsubsection{Soliton-soliton states}
The considerations for two-soliton states are analogous. Classically
the parameter $\mu$ characterizing the confining potential \fr{Vconf}
is the same as in the soliton-antisoliton sector, but we don't expect
this to be true at the quantum level. We account for this by a
different strength $\bar{\mu}$ of the potential, which then gives a
sequence of energies 
\be
E^{(T)}_n=2M_s+\left(\frac{4\bar{\mu}^2}{M_s}\right)^\frac{1}{3}\xi_n\ ,\quad
{\rm Ai}(-\xi_n)=0.
\label{Etrans}
\ee
Here we have taken the wave function to be antisymmetric because the
zero-momentum limit of the soliton-soliton scattering matrix is $-1$.
As soliton-soliton states have $S^z=1$, the bound states \fr{Etrans}
will be visible in the transverse structure factor.
\subsubsection{Dynamical Structure Factor at $Q\approx\pi/a_0$}
In the field theory limit the staggered magnetizations are given by
\be
\boldsymbol{n}(t,x)=
\begin{pmatrix}
\cos\big(\sqrt{2\pi}\Theta(t,x)\big)\\
\sin\big(\sqrt{2\pi}\Theta(t,x)\big)\\
\sin\big(\sqrt{2\pi}\Phi(t,x)\big)
\end{pmatrix}.
\ee
Close to the antiferromagnetic wave number $\pi/a_0$ (where $a_0$ is
the lattice spacing) the components of the DSF are thus given by
\be
S^{\alpha\alpha}\big(\omega,\frac{\pi}{a_0}+q\big)\propto\int_{-\infty}^\infty dt
dx\ e^{i\omega t-iqx}
\langle n^\alpha(t,x)n^\alpha(0,0)\rangle.
\ee
In the longitudinal structure factor we therefore see confined
soliton-antisoliton states, while the transverse components are
sensitive to confined soliton-soliton and antisoliton-antisoliton
states. Following Appendix B of \cite{DM} we can derive expressions
for the bound state contributions to the various correlators in
leading order in perturbation theory in $\mu$ in the limit of very
weak confinement
\bea
\langle n^z(\tau,x)\big)n^z(0,0)\big)\rangle&\propto&\sum_{n=1}^{N_b}
\big(E^{(L)}_{n+1}-E^{(L)}_n\big)\ \rho\big(E^{(L)}_n\big)\nn 
&&\times K_0(E^{(L)}_n|x^2+v^2\tau^2|),
\eea
where
\be
\rho(E)=\frac{2f\big({\rm arccosh}(E/2M_s)\big)}
{\pi^2\sqrt{E^2-4M_s^2}}\ .
\ee
The function $f(\theta)$ is related to a particular two-particle form factor
\bea
&&f(\theta)=\Big|\langle
0|n^z(0,0)|\theta_1,\theta_2\rangle_{s\bar{s}}\Big|^2\Bigg|_{\theta_1-\theta_2=\theta}=
C\frac{\sinh^2(\theta)}{\sinh^2(3\theta/2)}\nn
&&\times
\exp\left[-2\int_0^\infty\frac{dx}{x}
\frac{[\cosh x \cos\big(\frac{x\theta}{\pi}\big) -1] \cosh\big(\frac{x}{6}\big)}
{\cosh\big(\frac{x}{2}\big)\ \sinh x}\right].
\label{rho}
\eea
A similar analysis can in principle also be carried out on the level
of the spin chain itself. The strengths of the confining potentials
for spinon-antispinon and spinon-spinon two-particle states should
be extracted from the known 4-particle form factor. 

\subsection{DMRG Results}
\begin{figure}
\begin{center}
\includegraphics[trim=0cm 0cm 0cm 0cm, width=0.95\linewidth,clip]{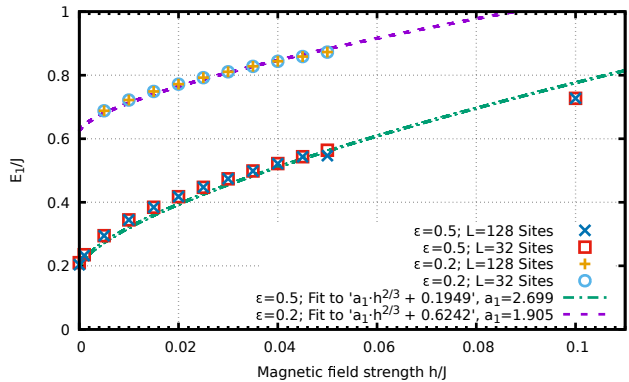}
\epsfxsize=0.43\textwidth
\end{center}
\caption{(Color online) Energy gap $E_1$ of the first bound state as a function of the magnetic field for $\eps =0.5$ and $\eps=0.2$. The lines are fits to the functional form \fr{gapfit}, where the zero field gap has been computed from the exact solution and $a_1=2.699$ and $a_1=0.6242$ for $\eps = 0.5$ and $\eps=0.2$ respectively. The finite-size effects between system sizes $L=32$ and $L=128$ are already too small to resolve graphically in most cases.}
\label{fig:dmrg1}
\end{figure}

While the strong coupling expansion and field theory analysis provide
a good qualitative picture of the dynamics, they do not apply
quantitatively to the experimentally relevant regime $\epsilon\approx
0.56$. In order to overcome this shortcoming we have carried out numerical DMRG \cite{White.PRL.69.2863, Schollwock.AnnPhys.326.96, Hubig.PRB.91.155115}
calculations with the \textsc{SyTen} tensor toolkit, based on the
Hamiltonian \eqref{HMF}. We first determine the gap of the lowest
bound state in the $S^z=1$ sector as the difference between the lowest
energies of the $S^z=0$ and $S^z=1$ sectors. This computation is
extremely stable, even near criticality at small values of the field
$h$ and with periodic boundary conditions. Finite-size effects can
also be entirely removed by choosing sufficiently large system
sizes. We are not able to determine the gaps of higher bound states
in this way as this would require working at a fixed momentum.
Fig.~\ref{fig:dmrg1} gives the resulting values for the gap and a fit
to the small-$h$ prediction \fr{gapfit}. We see that already for system
sizes $L=32$ the gap value is essentially converged and is in
excellent agreement with the theoretical predictions for the scaling
\fr{gapfit}. 

Second, we can calculate the ground state on a long chain, apply an
excitation in the middle of the chain and then use Matrix Product
State-based Krylov time evolution\cite{mcculloch-krylov} with matrix
re-orthogonalization to evaluate the dynamical structure factors in
the time-space domain. The Fourier transformation into momentum space
is unproblematic. However, we are only able to evolve up to a time
$t_{\mathrm{max}}J \approx 80$ for $\eps = 0.2$. This limitation is a
consequence of the entanglement growth during time evolution and the
subsequent exponential increase in computational effort. This limit
necessitates an articial damping factor
$\mathrm{exp}\left(-\eta \frac{t}{t_{\mathrm{max}}}\right)$ to be
introduced during the Fourier transform into frequency space. For
$\eps=0.2$, $\eta \approx 1$ suffices and it is already possible to
distinguish the physical peaks from the spectral leakage introduced by
the transformation. For $\eps=0.5$, only slightly shorter time-scales
$t_{\mathrm{max}}J \approx 50$ are achievable. Sufficient damping to
remove spectral leakage then also removes the signal. To circumvent
this problem, we use numerical extrapolation prior to the Fourier
transformation to extend the data in time to very large
$t^\prime_{\mathrm{max}}J \approx 1000$. We can then introduce a very
small damping $\eta = 1/700$ during the Fourier transformation and
still remove all spectral leakage.

\begin{figure}[ht]
\begin{center}
\includegraphics[width=0.95\linewidth,clip]{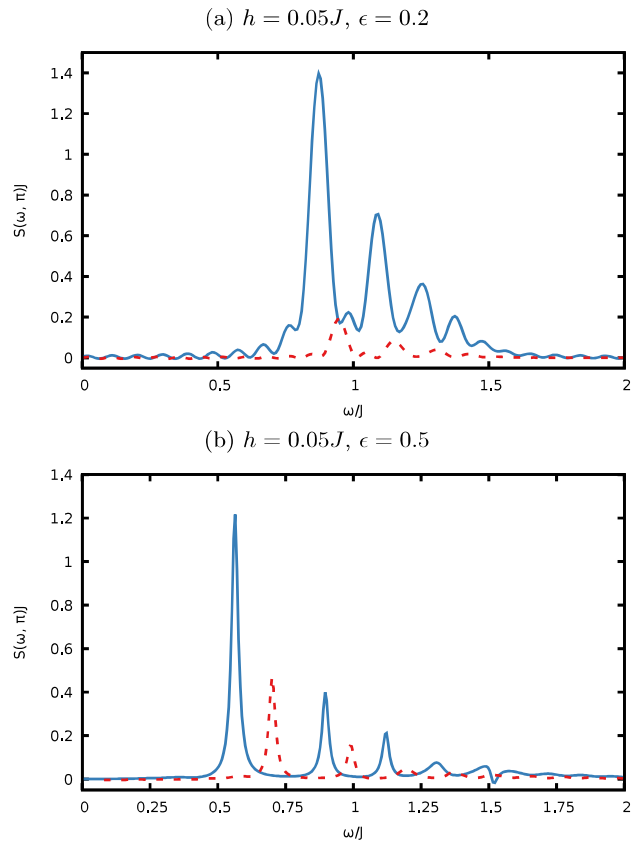}
\end{center}
\caption{(Color online) Transverse (solid blue line) and longitudinal (dashed red line) dynamical structure factors calculated using MPS-Krylov at (a) $h=0.05J$, $\eps=0.2$ and (b) $h=0.05J$, $\eps=0.5$. For $\eps=0.2$. 
The first peak for $\eps$=0.2 and 0.5 lie at $\omega = 0.87J$ and 0.56$J$, respectively, which are in good agreement with the value $E_1(h=0.05J) = 0.87J$ and 0.54$J$, respectively, extracted from Fig.~\ref{fig:dmrg1}. Some spurious oscillations appear in (a) is spite of the damping employed in the Fourier transformation. In (b), as a result of the numerical extrapolation procedure employed to deal with the late time regime, no spurious oscillations are visible.}
\label{fig:dmrg2}
\end{figure}

In Figs~\ref{fig:dmrg2} we show results for the
dynamical structure factor at momentum $Q=\pi$ on a system of $L=128$
sites with $h=0.05J$ and two values of the anisotropy $\eps$. In both
cases we find two sequences of bound states associated with the
transverse and longitudinal correlations respectively. The positions
of the first peak in the transverse sector is in good agreement with
Fig.~\ref{fig:dmrg1}.

\subsection{Tangent-space MPS methods}

\par As we have explained in the previous section, targeting higher bound states variationally requires the ability to work within a fixed momentum sector. This is made possible by using tangent-space methods \cite{Haegeman2013b} for matrix product states (MPS) that work directly in the thermodynamic limit. In particular, starting from a translationally-invariant MPS ground state on an infinite chain, we can apply the MPS quasiparticle ansatz \cite{Haegeman2012a} to target the elementary excitations corresponding to isolated branches in the spectrum. This ansatz can be read as the MPS version of the Feynman-Bijl ansatz and single-mode approximation, but improves on these approaches in using the virtual degrees of freedom of the MPS to build an excitation on top of the ground state. As the ansatz explicitly contains a fixed momentum, it allows to systematically capture the wave functions of all quasiparticle excitations -- the ones that contribute a $\delta$ peak in the DSF -- within a certain momentum sector. In order to capture continuous bands in the spectrum, multi-particle excitations should be considered \cite{Vanderstraeten2015a, Vanderstraeten2016}. Since the quasiparticle ansatz yields accurate variational expressions for the wave functions of the excited states, we can compute the energies and spectral weights for all states contributing to the DSF.

\par As it works directly in the momentum-energy plane, and does not
suffer from finite-size effects, this method has access to the
dynamical structure factor with perfect resolution. The only source of
error is the variational nature of the approach, but the approximation
can be systematically improved by growing the bond dimension of the
MPS ground state. As the ansatz effectively exploits the correlations
in the ground state to build an excitation, it can treat generic
strongly-correlated spin chains with isolated branches in the spectrum
to very high precision. An assessment of the accuracy of the
variational wave function for a given excited state is provided by the
variance $\langle H^2\rangle-\langle H\rangle^2$ of its energy, which can be evaluated exactly \cite{Vanderstraeten2015a}.

\par
\begin{figure}[ht]
\begin{center}
\epsfxsize=0.45\textwidth
\includegraphics[trim=0cm 0cm 0cm 0cm,clip=true, width=0.95\linewidth]{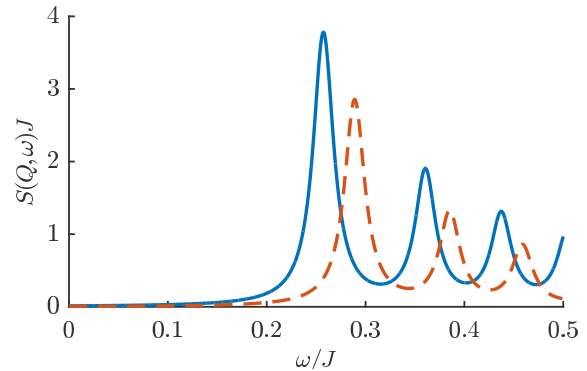}
\end{center}
\caption{(Color online) Transverse (solid blue lone) and longitudinal (dashed red line) structure factors for $J=7{\rm meV}$, $\epsilon=0.56$ and $h=0.00643J$. A Lorentzian broadening with width $\eta=J/80$ has been introduced to make the delta-function peaks visible.}
\label{fig:MPS}
\end{figure}

 At small values of the magnetic field $h$, we have a large number of stable bound states that live on a strongly-correlated background. Whereas time-domain approaches are necessarily limited in resolving the different modes, the quasiparticle ansatz is ideally suited for capturing all stable bound states with perfect resolution. Targeting the unstable bound states in the continuous bands would require a multi-particle ansatz \cite{Vanderstraeten2016}, but this has not proven to be necessary here. In order to compare with the experimental data, we have determined the dynamical structure factor by this method for several values of the anisotropy $\epsilon$ and the staggered magnetic field $h$. An additional shift of the energies has been introduced, corresponding to the three-dimensional dispersion of the modes. The best agreement with the experimental data is found for $\epsilon=0.56$ and $h=0.00643J$, and the corresponding structure factors are shown in Fig.~\ref{fig:MPS}.

\subsubsection{Comparison with strong-coupling expansion and DMRG results}
\begin{figure}
\begin{center}
\includegraphics[trim=0cm 0cm 0cm 0cm,clip=true, width=0.95\linewidth]{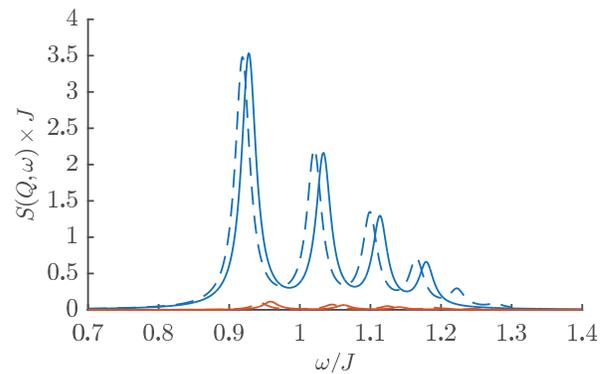}
\end{center}
\caption{(Color online) Transverse (blue lines) and longitudinal (red lines) structure factors for $J=7{\rm meV}$, $\epsilon=0.1$ and $h=0.025J$. The results of the MPS quasiparticle ansatz (solid lines) and the strong coupling approach (dashed lines) are in good agreement.}  
\label{fig:MPSstrong}
\end{figure}

It is useful to compare the results obtained by our different methods. We first consider a fairly strong anisotropy $\epsilon=0.1$ and weak field $h=0.025J$. Results for the quasiparticle ansatz (solid line) and the strong coupling approach (dashed line) are shown in Fig.~\ref{fig:MPSstrong}. The agreement is seen to be good and any discrepancies can be attributed to the leading ${\cal O}(\epsilon^2)$ corrections to the strong coupling result.

\begin{figure}[ht]
\begin{center}
\epsfxsize=0.45\textwidth
\includegraphics[trim=0cm 0cm 0cm 0cm,clip=true, width=0.95\linewidth]{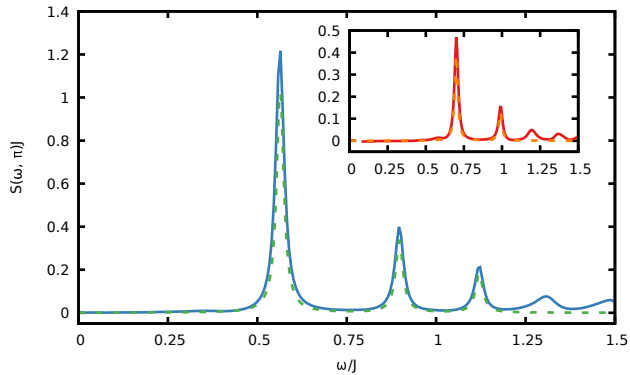}
\end{center}
\caption{(Color online) Transverse (blues lines, main figure) and longitudinal (red lines, inset) structure factors for $J=7{\rm  meV}$, $\epsilon=0.5$ and $h=0.05J$ calculated by DMRG (solid lines)
  and tangent-space MPS methods (dashed lines). For the chosen
  parameter set the agreement is seen to be excellent.}  
\label{fig:MPSdmrg}
\end{figure}

We have also compared the results of the quasi-particle ansatz to DMRG
for $\epsilon=0.5$, \emph{cf.} Fig.~\ref{fig:MPSdmrg}. The two methods
are seen to agree very well. The remaining differences arise from the
fact that the tangent-space MPS approach has been restricted to the
calculation of the five lowest energy bound modes (in principle higher
bound modes could be analyzed as well).

\subsubsection{Temperature effects}
\label{ssec:tempeff}
So far our theoretical analysis has been restricted to zero
temperature. In order to access the $T>0$ regime we now combine the
tangent-space MPS method with a low temperature linked cluster
expansion of the dynamical
susceptibility \cite{EK08,EK09,James.PhysRevB.78.094411,Goetze.Phys.Rev.B.82.104417}.
The basic idea is to treat the low-temperature regime as a gas of
bound states that scatter purely elastically. This is expected to be
a good approximation as long as the temperature is small compared to
the minimal gap $\Delta_{\rm min}$ of the lowest energy bound state, i.e.
\be
\exp\left(-\frac{\Delta_{\rm min}}{k_BT}\right)\ll 1.
\ee
The main temperature effects are a broadening of the $T=0$ coherent
single-particle peaks and the emergence of additional peaks in the
dynamical structure factor, which correspond to transitions between
thermally populated single-particle excitations. The first effect
requires an analysis of matrix elements between single particle and
two particle excitations. This is a non-trivial task beyond the scope
of the present work. At sufficiently low frequencies
$\omega<\Delta_{\rm min}$ the second effect is easier to capture. Let
us denote the single-particle excitations of the $a^{\rm th}$ bound
state with momentum $p$ by
\be
|p\rangle_a\ ,\quad p\in[0,\pi],
\ee
and the corresponding dispersion relations by $\epsilon_a(p)$.
Then the leading contributions to the dynamical structure factor at
low temperatures and frequencies are
\begin{widetext}
\bea
S^{\alpha\alpha}(\omega,Q)\Bigg|_{0<\omega,T<\Delta_{\rm min}}&\approx&
\frac{1}{1-e^{-\omega/k_BT}}
\sum_{a,b}\int_0^\pi \frac{dp}{\pi}\left[
e^{-\epsilon_a(p)/k_BT}-e^{-\epsilon_b(p+Q)/k_BT}\right]
\delta\big(\omega+\epsilon_a(p)-\epsilon_b(p+Q)\big)
{\cal M}_{ab}(p,Q)\ ,\nn
{\cal M}_{ab}(p,Q)&=&\left|{}_a\langle
p|S^\alpha_0|p+Q\rangle_b\right|^2
+\left|{}_a\langle p|S^\alpha_1|p+Q\rangle_b\right|^2
+2{\rm Re}\left[e^{-iQ}{}_a\langle p|S^\alpha_1|p+q\rangle_b
{}_b\langle p+Q|S^\alpha_0|p\rangle_a\right].
\label{lowT}
\eea
\end{widetext}

\begin{figure}[ht]
\begin{center}
\includegraphics[trim=0cm 0cm 0cm 0cm,clip=true, width=0.95\linewidth]{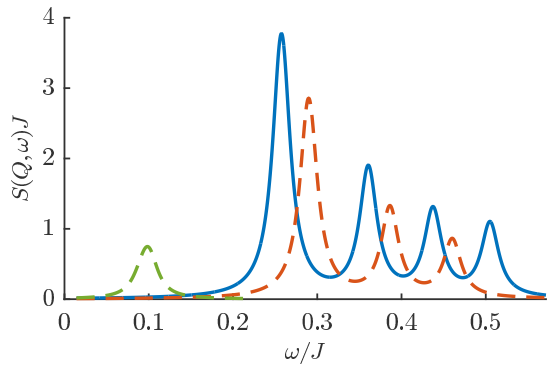}
\end{center}
\caption{Dynamical structure factor calculated with tangent-space MPS
methods for $\eps=0.56$, $J=7{\rm meV}$, $h=0.00614J$. Finite
temperature resonances given by \fr{lowT} have been determined at
temperature $T=4K$ and are plotted together with the $T=0$ results for
the higher energy bound modes to allow for a comparison of the
relative intensities. Longitudinal (transverse) components are plotted
with dashed (solid) lines. A Lorentzian broadening with width $\eta=J/80$ has been introduced to make the delta-function peaks visible. } 
\label{fig:MPST}
\end{figure}

In Fig.~\ref{fig:MPST} we show the contributions \fr{lowT} due to
transitions between thermally excited bound modes for the
experimentally relevant parameter set $\eps=0.56$, $J=7{\rm meV}$,
$h=0.00614J$ and $T=4K$. Transitions occurring at
very low frequencies have not been taken into account, because the low
energy regime is dominated by the broadened Bragg peak. For comparison
the longitudinal (red dashed line) and transverse (solid blue line)
components of the dynamical structure factor at $T=0$ are shown as
well. A finite temperature resonance in the longitudinal structure
factor at a frequency $\omega\approx 0.1J$ is clearly visible, while
contributions to the transverse structure factor are very small.

\section{Theory vs experiment}
We are now in a position to compare theoretical and experimental
results. The first task is to determine appropriate parameters for
applying the effective 1D model \fr{HMF} to \SrCoVO. Estimates for the
exchange $J$ and anisotropy $\epsilon$ were obtained in section
\ref{sec:HighT} by comparing the data collected for $T>T_N$ to the
zero temperature transverse dynamical structure factor for \fr{HMF}
with $h=0$. Such a comparison is appropriate because $T\ll J$ and
gives values of $J\approx 7.0$~meV and $\epsilon=0.56$. The
remaining parameter is the strength $h$ of the effective staggered
field. As this arises from a mean-field decoupling of the interchain
interactions, it is temperature dependent. As shown in section
\ref{sec:theory}, $h$ can be fixed by computing the energies of the
first few bound states and comparing them to the measured peak
positions. One caveat is that the gap of the lowest bound state is not
necessarily well described by the purely 1D model \fr{HMF}. Indeed, in
simple quasi-1D systems of weakly coupled chains corrections to the
simple mean-field approximation due to the interchain couplings can be
taken into account by a random-phase approximation, which gives the
following expression for the dynamical susceptibility
\be
\chi_{\rm 3D}(\omega,{\bf q})=\frac{1}
{\chi^{-1}_{\rm 1D}(\omega,q)-J_{\rm int}({\bf q})}\ .
\label{RPA}
\ee
Here $J_{\rm int}({\bf q})$ is the Fourier transform of the interchain
coupling, and we have assumed that we are dealing with a system of
equivalent chains. It is clear from \fr{RPA} that at a given wave
vector the singularities of $\chi_{\rm 3D}(\omega,{\bf q})$ are
shifted in energy by a constant compared to those of $\chi_{\rm
  1D}(\omega,q)$. The situation in \SrCoVO\ is much more complicated,
because there are several counter rotating screw chains per unit cell. A refined
analysis of the interchain coupling by a generalization of \fr{RPA} is
possible\cite{Zheludev} but beyond the scope of this work.

\begin{figure}
\begin{center}
\includegraphics[trim=0cm 0cm 0cm 0cm,clip=true, width=0.95\linewidth]{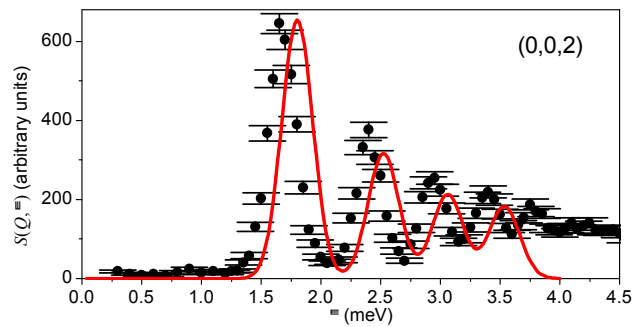}
\end{center}
\caption{(Color online) Transverse dynamical structure factor calculated with tangent-space MPS methods for $\eps=0.56$, $J=7$~meV, $h=0.00614J$ (solid line) compared to experimental data for $T=1.5$~K and wave vector transfer $(0,0,2)$. An artificial Gaussian broadening with width $0.26$~meV was introduced to mimic the experimental resolution. The MPS results were rescaled in order to match the amplitude of the first peak. The MPS computation was restricted to determining the first four bound modes only.} 
\label{fig:MPSexpt}
\end{figure}

Keeping this discussion in mind, we first try to obtain an optimal
description of the energy splittings between the observed coherent
modes by a pure one dimensional model. At temperature $T=1.5$~K we can
reproduce the energy differences of the first few peaks with a value
of $h\approx 0.00643J$. The resulting comparison between the transverse modes calculated by this mean-field model using the tangent-space MPS method \fr{HMF} and the experimental data at $Q=(0,0,2)$ is shown in Fig.~\ref{fig:MPSexpt}. 
We see that the mean-field model reproduces the experimental results
very well up to an overall shift of about $0.122{\rm meV}$ in energy.
Since the gap is very sensitive to corrections to the mean-field
model, as can be seen from the RPA expression \fr{RPA}, such a shift
is not surprising. Furthermore in the experimental data the interchain 
couplings give rise to a dispersion of the gap with a comparable bandwidth 
of $\sim 0.15$~meV (see Fig.~\ref{Fig:perp_dispersion}).

\begin{figure}[ht]
\includegraphics[trim=0cm 0cm 0cm 0cm,clip=true, width=0.95\linewidth]{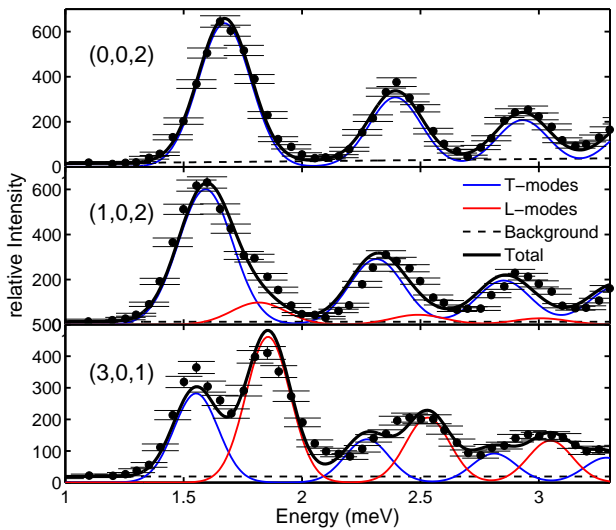}
\caption{\label{fig:MPSexpt_2}(Color online) 
The transverse (blue line) and longitudinal (red line) dynamical structure factors calculated by the tangent-space MPS method for $\eps=0.56$, $J=7{\rm meV}$, $h=0.00614J$ compared to the experimental data (filled circles) for $T=1.5K$ measured at wavevector transfers (a) (0,0,2), (b) (1,0,2) and (c) (3,0,1). At each wavevector the intensity of the two structure factors are weighted by their respective polarization factors and by the square of their $g$-factors \cite{Zhe.PRB.94.125130} giving a weighting ratio of longitudinal to transverse modes of 0, 0.216, 2.16 for (0,0,2), (1,0,2) and (3,0,1) respectively; an overall scale factor is also included to match the data. To account for the effects of interchain coupling a wavevector-dependent energy shift is introduced with values -0.13, -0.21, -0.25 meV for the three wavector respectively. Finally the theoretical peaks are convolved by Gaussians of widths 0.26, 0.27 and 0.22meV to model the experimental resolution. The two structure factors are summed together along with a linear background (dashed black line) to give the expected total scattering (solid black line).} 
\end{figure}

The dynamical structure factor calculated in the mean-field model by the tangent-space MPS method was compared to the data at several reciprocal lattice points as shown in Fig.~\ref{fig:MPSexpt_2}. Both the transverse and longitudinal structure factors are plotted and the effect of interchain coupling is taken into account by introducing a wavevector-dependent energy shift. At each wavevector the intensity of the two structure factors are weighted by their respective polarization factors due to the component of their magnetization perpendicular to the wavevector transfer (see section \ref{ssec:lowTP}) as well as by the square of their $g$-factors \cite{Zhe.PRB.94.125130}. An overall scaling factor is also introduced to match the theoretical intensity to the data and the theoretical peaks are convolved by a Gaussian to model the experimental resolution. The solid black line gives the sum of the two structure factors as well as a linear background and represents the expected neutron scattering intensities. Considering the highly complex counter-rotating screw chain structure and the many possible interchain interactions that are neglected in this calculation, the agreement between experiment and theory is remarkably good.

\begin{figure}
\includegraphics[trim=0cm 0cm 0cm 0cm,clip=true, width=0.95\linewidth]{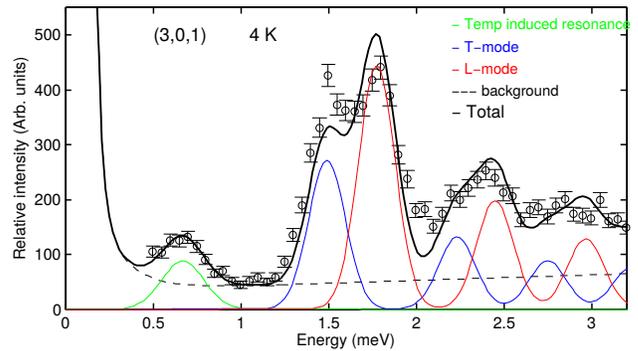}
\caption{\label{fig:boundstate_4K} (Color online) A comparison of the dynamical structure factors for the transverse (blue line) and longitudinal (red line) and the temperature induced resonance (green line) calculated by the tangent-space MPS method (Fig. \ref{fig:MPST}) with the experimental data (open circles) at $T=4$ K and measured at wavevector transfers (3,0,1). The temperature induced resonance (green line) given by \fr{lowT} has been determined at temperature $T=4K$, whereas, the dynamical structure factors for the T-mode and L-mode are calculated at $T$ =0 K. The relative intensities for the T-mode, L-mode, and temperature induced resonance peak are taken to be 1, 2.16 (same as  Fig.\ref{fig:MPSexpt_2}), and 0.325, respectively. An overall scale factor is also included to match the data. For L-mode and T-mode energy shifts of -0.31 and -0.25 meV, respectively, are introduced to account for the effects of interchain coupling. Finally the theoretical peaks are convolved by Gaussians of widths 0.24 for the temperature induced peak and 0.22 meV for the L- and T-modes to model the experimental resolution. The all structure factors are summed together along with a background (dashed black line) to give the expected total scattering (solid black line). The background was calculated by a combination of two functions; a linear and an exponential decay for which the coefficients are derived from the fitting of the 4.75 K experimental data for wavevector transfer (0,0,2) [Fig. \ref{Fig:temp_boundstates}(a)].}
\end{figure}

The finite temperature results of section \ref{ssec:tempeff} are also in good agreement with the experimental observations [Fig. \ref{fig:boundstate_4K}]. We saw that at a temperature of $4$~K our one dimensional model displays a finite temperature resonance in the longitudinal structure factor at an energy of about $0.7$~meV. This is in good agreement with the experimental observation of a resonance in the longitudinal structure factor at $\omega\approx 0.65$~meV [Fig.~\ref{Fig:temp_boundstates}(b)]. 

\section{Discussion}
We have presented results of inelastic neutron scattering experiments on the quasi-one dimensional spin-1/2 Heisenberg magnet \SrCoVO. Above the N\'eel temperature $T_N\approx 5.2K$, the neutron scattering cross section is dominated by a scattering continuum that is well described by a spin-1/2 Heisenberg XXZ chain with antiferromagnetic exchange $J\approx 7.0\pm 0.2\ {\rm meV}$ and anisotropy parameter
$\epsilon\approx 0.56\pm 0.02$. The scattering continuum is formed by fractionalized $S^z=\frac{1}{2}$ spinon excitations. At temperatures below $T_N$ the structure factor exhibits two sequences of
resolution-limited dispersing peaks that are associated with fluctuations along (L-modes) and perpendicular (T-modes) to the ordered magnetic moment respectively. 


The origin of these coherent modes can be understood by a one dimensional model [Eq. \fr{HMF}], in which a (temperature dependent) staggered magnetic field is generated in the ordered phase through a
mean-field decoupling of the interchain interactions. The model can be studied analytically for both strong and weak exchange anisotropies and in both limits the effect of the staggered field is to confine the spinon excitations into two sequences of bound states. At intermediate values of the exchange anisotropy we have used DMRG and MPS methods to obtain quantitative results for the dynamical structure factor. It turns out that the experimentally relevant
parameter regime cannot be reached even by state-of-the-art DMRG
methods. Due to entanglement growth, the time scale by which dynamical
correlation functions that can be computed by DMRG is restricted, which in
turn imposes limitations on the achievable energy resolution. We
therefore have employed a recently developed tangent-space MPS
method, which is based on constructing MPS representations for excited
states. Application of this method allows the computation of the
dynamical structure factor, which is found to be in good agreement
with experiment. 

Our work establishes \SrCoVO\ as a beautiful paradigm for spinon
confinement in a quasi-one dimensional quantum magnet. There are a
number of interesting questions that deserve further
investigation. On the theoretical side a more involved investigation of the dynamical structure factor at finite temperatures would improve our understanding of the thermally induced peaks observed in the data both below $T_N$ (transitions between bound modes) and above $T_N$ (the villain mode). On the experimental side, the precise form of the
interchain interactions needs to be clarified by extensive measurements of the bound mode dispersion relations perpendicular to the chain direction at lowest temperatures. We have seen that it is
necessary to account for these interations beyond a simple mean-field decoupling in
order to describe the data. As the crystal structure is rather
complex this goes beyond the scope of the present work. Finally, it
would be interesting to analyze the effects of an applied uniform
magnetic field. Terahertz spectroscopy measurements reveal the emergence of novel excitations as a function of both transverse and longitudinal magnetic field \cite{Zhe.PRB.94.125130,Zhe_Long_Field} which could be investigated using a combination of neutron scattering and the theoretical methods described here. We hope to return to these questions in future work.\\

\acknowledgments
We acknowledge the Helmholtz Gemeinschaft for 
funding via the Helmholtz Virtual Institute (Project No. HVI-521). This work was supported by the EPSRC under grant EP/N01930X/1 (FHLE), the ExQM graduate school and the Nanosystems Initiative Munich (CH). L. Vanderstaeten acknowledge the financial support from FWO Flanders.

\end{document}